\documentclass[structabstract]{aa}  
\usepackage{graphicx}
\usepackage{txfonts}
\begin{document}
   \title{Molecular content of the circumstellar disk in AB~Aur}
   \subtitle{First detection of SO in a circumstellar disk}

   \author{A. Fuente
          \inst{1}
	  \and
           J. Cernicharo
          \inst{2}
          \and
           M. Ag\'undez
          \inst{3}
	  \and
          O. Bern\'e
          \inst{4}
          \and
          J. R. Goicoechea
          \inst{2}
          \and
          T. Alonso-Albi
          \inst{1}
          \and
          N. Marcelino
          \inst{2}
}
   \institute{ Observatorio Astron\'omico Nacional (OAN), Apdo. 112,
             28803 Alcal\'a de Henares, Madrid, Spain
        \and
            Departamento de Astrof\'{\i}sica, Centro de Astrobiolog\'{\i}a (CSIC-INTA), Crta Ajalvir km 4, 28850 Madrid, Spain 
         \and
            LUTH, Observatoire de Paris-Meudon, 5 place Jules Janssen 92190 Meudon, France
          \and
        Leiden Observatory, Leiden University, PO Box 9513, 2300 RA Leiden, The Netherlands
}
 \abstract 
   {}
   {Very few molecular species have been detected in circumstellar disks surrounding young stellar objects. We are carrying out an observational study of the chemistry of circumstellar disks surrounding T Tauri and Herbig Ae stars. First results of this study are presented in this note.}
   {We used the EMIR receivers recently installed at the IRAM 30m telescope to carry a sensitive search for molecular lines in the disks surrounding 
AB~Aur, DM~Tau, and LkCa~15.}
   {We detected lines of the molecules HCO$^+$, CN, H$_2$CO, SO, CS, and HCN toward AB Aur. In addition, we tentatively 
detected DCO$^+$ and H$_2$S lines. The line profiles suggest that the CN, HCN, H$_2$CO, CS and SO lines arise in the disk. This makes it the first 
detection of SO in a circumstellar disk. We have
unsuccessfully searched for SO toward DM Tau and LkCa 15, and for c-C$_3$H$_2$ toward AB~Aur, DM~Tau, and LkCa~15. Our upper limits 
show that contrary to all the molecular species observed so far, SO is not as abundant in DM Tau as it is in AB~Aur.}
 {Our results demonstrate that the disk associated with AB Aur is rich in molecular species. Our chemical model shows that the detection of SO is consistent with that expected from a very young disk where the molecular adsorption onto grains does not yet dominate the chemistry.
}
   \maketitle
%

\section{Introduction}
Circumstellar disks are complex systems in which essentially all the processes that play 
a role in the interstellar medium, UV radiation, X-rays, grain surface
chemistry, molecular depletion, turbulent mixing, accretion flows and time dependency,
are working. Chemical models with increasing complexity have been developed in the last decade
(see e.g. Aikawa et al. 2000; Dutrey et al. 2007; Ag\'undez, Cernicharo
\& Goicoechea, 2008; Nomura et al. 2009), but the disk chemistry is a quite unexplored field
from the observational point of view.
Large millimeter telescopes have started to provide some insight into the chemistry
of the cold gas toward the most massive nearby disks.
Thus far, few molecules (CO, $^{13}$CO, CN, C$_2$H, HCN, HNC, HCO$^+$, H$_2$CO) have been detected
in circumstellar disks.
This small molecular inventory is mainly due to the weakness of the molecular emission from circumstellar disks.
Disks have small masses, lower than 0.1~M$_\odot$, small sizes, radii of a few 100~AU, and because of 
depletion in the midplane and/or photodissociation in the surface, the disk averaged abundances of most molecules (including CO and its 
isotopologues) are a factor of 5--10 lower than in the interstellar medium. High sensitivity is, therefore, required for an observational
study. We have carried out a sensitive search for molecular lines mainly
in the disk around the Herbig Ae star AB~Aur using the IRAM 30m telescope. 
Some lines have also been searched toward DM~Tau and LkCa~15. Our results show the rich molecular
content in the disk around AB~Aur.

AB Auriga is one of the nearest, brightest and best studied Herbig Ae stars. It has a spectral type A0--A1 (Hern\'andez et al. 2004) and is located 
to the Southwest of the molecular cloud L1517 (Duvert et al. 1986),
at a distance of 145~pc (van den Ancker et al. 1998).
Interferometric observations at millimeter wavelengths detected the circumstellar disk around this star in the continuum and
in the CO (and its isotopologues) lines (Pi\'etu et al. 2005). 
Instead of being centrally peaked, the continuum  emission is dominated by a 
bright, asymmetric (spiral-like) feature at about 140~AU from the central star.

The disk modeling of the continuum and molecular emission showed that the disk is warm and showed no evidence of CO depletion. 
Schreyer et al. (2008) searched for emission of the HCO$^+$ 1$\rightarrow$0, HCN 1$\rightarrow$0, CS 2$\rightarrow$1, C$_2$H 1$\rightarrow$0 and 
some CH$_3$OH lines in this disk 
using the Plateau de Bure Interferometer (PdBI), but they only detected the HCO$^+$ 1$\rightarrow$0 line. 

\section{Observations}
The list of observed lines and the telescope parameters are given in Table~1. The observations were done in two observing periods, September, 2009 and March, 2010, with the new EMIR receivers arranged to provide a bandwidth of 4~GHz in both, the 3mm and 1mm bands. As backends we used the wide bandwidth autocorrelator WILMA which provides a spectral resolution of 2~MHz and covers the whole band, and the narrow bandwidth correlator VESPA centered at the line frequency and providing a spectral resolution of 80~kHz at 1.3mm and 40~kHz at 2.7mm ($\sim$0.1~km~s$^{-1}$).
All the observations were done using the wobbler switching (WS) procedure with a throw of 120$"$. This procedure provides flat baselines which are essential for detecting weak and wide lines toward compact sources, which is the case for the lines arising in circumstellar disks.
In the case of AB~Aur, the disk is still immersed in the parent cloud whose emission extends farther than the wobbler throw (see Semenov et al. 2005).
Then, at the velocities of the ambient cloud the detected emission is just the ON-OFF balance without any physical interpretation (remind that the OFF position is moving in the sky during the source tracking). For this reason, we have blanked the channels corresponding to the ambient cloud emission in 
the spectra toward AB~Aur.
We have searched for c-C$_3$H$_2$ and SO also toward DM~Tau and LkCa15.
In these cases, contamination from the ambient cloud is not expected.
The observational results are shown in Table 2.

\begin{table}
\begin{tabular}{lllll}\\
\multicolumn{5}{l}{ Table 1: List of targeted lines} \\ \hline \hline
\multicolumn{2}{c}{Line}     & \multicolumn{1}{c}{Freq.(GHz)} &
\multicolumn{1}{c}{HPBW($"$)} & \multicolumn{1}{c}{$\eta_b$}  \\   \\ \hline
CO       &  2$\rightarrow$1   &  230.538   &  10    &  0.63 \\
HCO$^+$  &  1$\rightarrow$0   &   89.188   &  28    &  0.81   \\
HCO$^+$$^1$  &  3$\rightarrow$2   &  267.558   &   9    &  0.53 \\
HCN      &  1$\rightarrow$0   &   88.631   &  28    &  0.81  \\
HCN      &  3$\rightarrow$2   &  265.886   &  9     &  0.53  \\
CN       &  1$\rightarrow$0   &  113.490   &  22    &  0.81  \\
CN       &  2$\rightarrow$1   &  226.874   &  10    &  0.63  \\
CS       &  2$\rightarrow$1   &   97.981   &  25    &  0.81  \\
CS       &  3$\rightarrow$2   &  146.969   &  16    &  0.74  \\
C$_2$H   &  3$\rightarrow$2   &  262.004   &   9    &  0.53 \\
H$_2$CO  &  3$_{0,3}$$\rightarrow$2$_{0,2}$    &  218.222   &  11    & 0.63  \\
SO  & 3$_4$$\rightarrow$2$_3$ &  138.178   &  17    & 0.74  \\
SO$^2$  & 5$_6$$\rightarrow$4$_5$ & 219.949   &  11    & 0.63 \\
c-C$_3$H$_2$    &  2$_{1,2}$$\rightarrow$1$_{0,1}$   &  85.339   & 29   & 0.81 \\
c-C$_3$H$_2$    &  6$_{0,6}$$\rightarrow$5$_{1,5}$   &  217.822  & 11   & 0.63 \\
c-C$_3$H$_2$    &  6$_{1,6}$$\rightarrow$5$_{0,5}$   &  217.822  & 11   & 0.63 \\
DCO$^+$       &  2$\rightarrow$1   &  144.077  & 16   & 0.74 \\
DCN           &  2$\rightarrow$1   &  144.828  & 16   & 0.74 \\
SiO           &  2$\rightarrow$1   &  86.847   & 29   & 0.81 \\
SiO           &  6$\rightarrow$5   &  260.518  & 9     &  0.53  \\
HCO           &  1$_{0,1} 3/2,2$$\rightarrow$0$_{0.0}$1/2,1    &  86.671   & 29   & 0.81 \\
HCO               &  1$_{0,1} 3/2,1$$\rightarrow$0$_{0.0}$1/2,0    &  86.708   & 29   & 0.81 \\
HCO               &  1$_{0,1} 1/2,1$$\rightarrow$0$_{0.0}$1/2,1    &  86.777   & 29   & 0.81 \\
HCO               &  1$_{0,1} 1/2,0$$\rightarrow$0$_{0.0}$1/2,1                   &  86.805   & 29   & 0.81 \\
H$_2$S        &  1$_{1,0}$$\rightarrow$1$_{0,1}$   & 168.763   &  14  & 0.74 \\
\hline \hline
\end{tabular}

\noindent
$^1$ Observed with the wide band spectrometer WILMA in the HCN 3$\rightarrow$2 tuning.

\noindent
$^2$ Observed with the wide band spectrometer WILMA in the c-C$_3$H$_2$ 6$_{0,6}$$\rightarrow$5$_{1,5}$ tuning.

\end{table}

\begin{figure}
\includegraphics{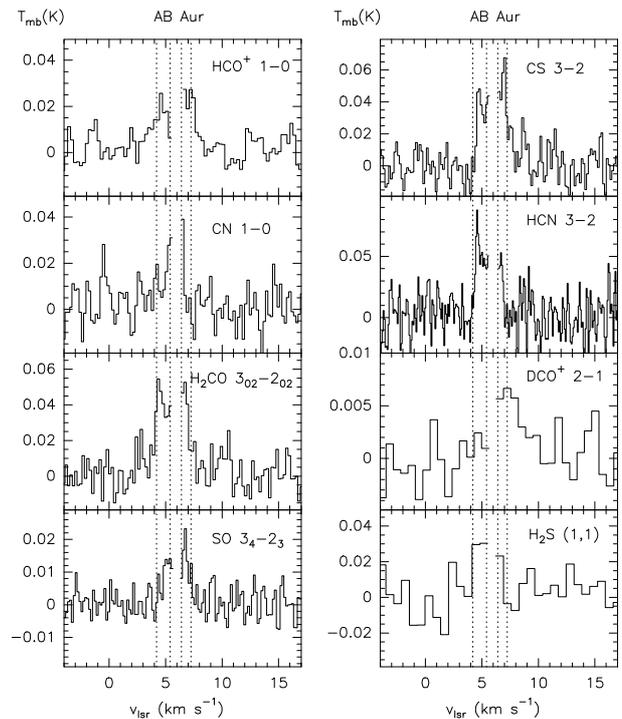}
\vspace{10.0cm}
      \caption{ Spectra obtained with the 30m telescope toward AB Aur. Dashed lines indicate the velocities at which the disk emission arises ([4.2,5.6] km~s$^{-1}$ and [6.5,7.25] km~s$^{-1}$ ). 
}
         \label{Fig 1}
 \end{figure}

\section{Results}

\begin{figure}
\includegraphics{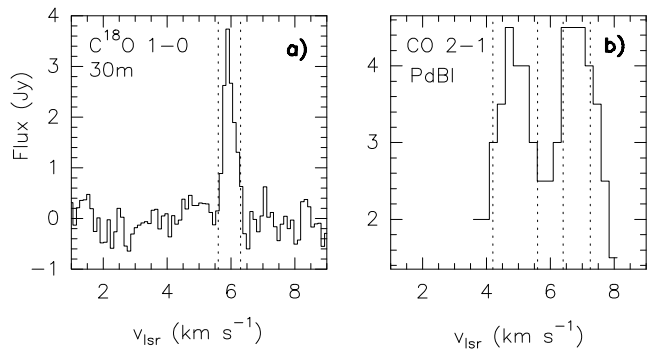}
\vspace{5.5cm}
      \caption{{\bf a)} Spectra of the C$^{18}$O 1$\rightarrow$0 toward AB~Aur observed with the IRAM 30m telescope by Fuente et al. (2002). We have
adopted this profile as a pattern profile for the ambient cloud emission. Vertical lines indicate the velocity interval [5.6,6.4] km~s$^{-1}$. {\bf b)} Interferometric spectra of 
the $^{12}$CO 2$\rightarrow$1 toward the star position (Fig. 2 (bottom) of Pi\'etu et al. 2005). Note that the disk emission occurs
at the velocity intervals, [4.2,5.6] km~s$^{-1}$ and [6.4,7.25] km~s$^{-1}$. These velocities are indicated by vertical dashed lines.
}
         \label{Fig 1}
\end{figure}

Fig.~1 shows some of the spectra observed toward AB~Aur.
The lines from the molecular cloud are very narrow, $\Delta v$$\sim$0.5~km~s$^{-1}$, and 
centered at 5.9~km~s$^{-1}$ (Duvert et al. 1986 and Fig.~2a). The emission of the ambient cloud 
lies at the velocities [5.4,6.4]~km~s$^{-1}$. The channels corresponding
to these velocities are blanked
in the spectra shown in Fig. 1.

After blanking the cloud velocities, we have detected emission at $>$3$\sigma$ of the HCO$^+$ 1$\rightarrow$0,
 CN 1$\rightarrow$0, H$_2$CO 3$_{0,3}$$\rightarrow$2$_{0,2}$, SO 3$_{4}$$\rightarrow$2$_{3}$, CS 3$\rightarrow$2, 
and HCN 3$\rightarrow$2 lines. In addition, we have tentatively detected ($\sim$3$\sigma$)
the DCO$^+$ 2$\rightarrow$1 and H$_2$S 1$_{1,0}$$\rightarrow$1$_{0,1}$ lines.
All the ($>$3$\sigma$) detected lines have the typical two-horn profile observed in the 
lines coming from the circumstellar disk,
with two peaks centered at 4.8$\pm$0.25~km~s$^{-1}$ and 6.8$\pm$0.25~km~s$^{-1}$ (see Fig.~1 and Fig.~2b).
This prompts us to interpret the emission of these lines as arising from the circumstellar disk.
The only doubtful case is the CN 1$\rightarrow$0 line in which the two-horn profile is not so clear.
Since CN is one of the most abundant species in disks (Dutrey et al. 1997, Thi et al. 2004, \"{O}berg et al. 2010),
we decided to keep it in our list of detected species.
The narrowness of the CN 1$\rightarrow$0 line could  be due to the fact that its emission is coming from the outermost part
of the disk. We would like to remind, however, that this detection requires of
further confirmation by interferometric observations.

The disk was previously detected in the HCO$^+$ 1$\rightarrow$0 line using the PdBI by Schreyer et al. (2008).
Therefore, we can use this line to check the validity of our interpretation.
In Fig. 3, we compare our HCO$^+$ spectrum with that observed 
toward the star position by Schreyer et al. (2008). Since
the synthesized beam of these observations was 5.2$"$ $\times$ 4.8$"$, this spectrum missed the emission
of the outer part of the disk (R$>$378~AU). The emission of this outer
region is expected at velocities $<$0.87~km~s$^{-1}$ from the systemic
velocity. We only consider velocities $>$0.45~km~s$^{-1}$ relative to
the systemic velocity, therefore the outer part of the disk is not relevant in our comparison. The integrated intensity emission of the 30m spectra
in the velocity intervals [4.2,5.4]~km~s$^{-1}$ and [6.4,7.25]~km~s$^{-1}$ is lower by a factor of $\approx$1.3 than the integrated emission of the HCO$^+$ 1$\rightarrow$0 line as observed with the PdBI (see Fig. 3). The agreement is acceptable and we consider that our interpretation is valid.

In Table 2, we show the list of non-detections (in the considered velocity ranges). Our 3$\sigma$ upper limit to 
the emission of CS 2$\rightarrow$1 line improve by a factor of 2 the previous one obtained by Schreyer et al. (2008)
using the PdBI. 

\begin{figure}
\includegraphics{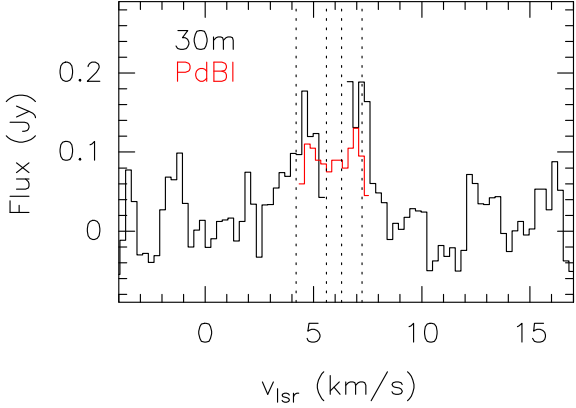}
\vspace{6.0cm}
      \caption{Comparison between the spectra of the HCO+ 1$\rightarrow$0 line observed toward AB~Aur with the 30m telescope and the PdBI. Dashed lines indicate the velocities at which the disk emission arises ([4.2,5.6] km~s$^{-1}$ and [6.5,7.25] km~s$^{-1}$ ). 
}
         \label{Fig 1}
 \end{figure}

\begin{table}
{\small
\begin{tabular}{llcc|llc}\\
\multicolumn{7}{l}{ Table 2: Observational results} \\ \hline \hline
\multicolumn{4}{c|}{ Detections}  & \multicolumn{3}{c}{Non-detections}   \\ 
\multicolumn{2}{c}{Line}     & \multicolumn{1}{c}{Area$^1$}  & \multicolumn{1}{c|}{rms$^2$} &
\multicolumn{2}{c}{Line}     & \multicolumn{1}{c} {rms$^2$}  \\ 
\multicolumn{3}{r}{(mK$\times$km s$^{-1}$)}  & \multicolumn{1}{c|}{(mK)}  &
\multicolumn{2}{c}{} & \multicolumn{1}{c}{(mK)} \\ \hline
\multicolumn{4}{c|}{AB~Aur}     & \multicolumn{3}{c}{AB~Aur} \\ 
HCO$^+$  &  1$\rightarrow$0   &    47       &  4                 & CN     &  2$\rightarrow$1  &  16    \\
HCO$^+$  &  3$\rightarrow$2   &   932$^3$   &  4$^3$             & CS     &  2$\rightarrow$1  &  4  \\ 
CN       &  1$\rightarrow$0   &    26       &  6                 & C$_2$H &  3$\rightarrow$2  &  7  \\
CS       &  3$\rightarrow$2   &    75       &  7                 & HCN   &  1$\rightarrow$0  &  6  \\ 
H$_2$CO  &  3$_{0,3}$$\rightarrow$2$_{0,2}$   &    87     &  5     & c-C$_3$H$_2$    &  2$\rightarrow$1 &      4  \\  
SO       &  3$_4$$\rightarrow$2$_3$           &    26     &  4     & c-C$_3$H$_2$    &  6$\rightarrow$5 &      4  \\ 
SO       &  5$_6$$\rightarrow$4$_5$           &    64$^3$ &  2$^3$ & DCN           &  2$\rightarrow$1   &      4  \\  
HCN      &  3$\rightarrow$2                   &    92     &  6     &  SiO           &  2$\rightarrow$1  &  4  \\  
DCO$^+$       &  2$\rightarrow$1              &     8      &  2           &  SiO           &  6$\rightarrow$5  &  9   \\
H$_2$S        &  1$_{1,0}$$\rightarrow$1$_{0,1}$   &  69   &  11   &  HCO           &  1$_{0,1}$$\rightarrow$0$_{0.0}$   &  4   \\ 
\multicolumn{4}{c|}{}   & \multicolumn{3}{c}{ DM Tau}   \\
\multicolumn{4}{c|}{}  &  c-C$_3$H$_2$    &  2$\rightarrow$1   &    3  \\
\multicolumn{4}{c|}{}  & c-C$_3$H$_2$    &  6$\rightarrow$5   & 6  \\ 
\multicolumn{4}{c|}{}  & SO            &  5$_6$$\rightarrow$4$_5$     & 3$^3$  \\
\multicolumn{4}{c|}{}  & \multicolumn{3}{c}{ LkCa15} \\
\multicolumn{4}{c|}{}  & c-C$_3$H$_2$    &  2$\rightarrow$1   &  3  \\
\multicolumn{4}{c|}{}  & c-C$_3$H$_2$    &  6$\rightarrow$5   &  9  \\ 
\multicolumn{4}{c|}{}  & SO            &  5$_6$$\rightarrow$4$_5$    &   3$^3$  \\
\hline \hline
\end{tabular}

\noindent
$^1$Sum of the integrated intensity area in the velocity intervals [4.2,5.6]+[6.4,7.25]~km~s$^{-1}$.

\noindent
$^2$rms in a channel of 1~km~s$^{-1}$.

\noindent
$^3$Observed only with a velocity resolution of 2.7~km~s$^{-1}$.
}
\end{table}

\begin{table*}
\begin{tabular}{lccccccc}\\
\multicolumn{8}{c}{Table 3: Comparison with other disks$^1$}  \\ \hline \hline
\multicolumn{1}{c}{}     & \multicolumn{2}{c}{AB Aur}  & \multicolumn{1}{c}{DM Tau}  & \multicolumn{1}{c}{LkCa15}  &  \multicolumn{1}{c}{TW~Hya} &  
\multicolumn{1}{c}{HD163296}  & \multicolumn{1}{c}{MWC~480} \\
\multicolumn{1}{c}{}     & \multicolumn{1}{c}{T$_{rot}$=10~K}   & \multicolumn{1}{c}{T$_{rot}$=20~K} &
\multicolumn{1}{c}{}     & \multicolumn{1}{c}{}                 & \multicolumn{1}{c}{}                &
 \multicolumn{1}{c}{}  &  \multicolumn{1}{c}{} \\  \hline \hline
X(HCO$^+$)$^1$    & 1.8$\times$10$^{-11}$  & 2.8$\times$10$^{-11}$  &  7.4$\times$10$^{-10}$  &  5.6$\times$10$^{-12}$  & 2.2$\times$10$^{-11}$    & 
 7.8$\times$10$^{-12}$   & 1.0$\times$10$^{-10}$ \\
X(CN)         & 1.0$\times$10$^{-10}$  & 1.4$\times$10$^{-10}$      &  3.2$\times$10$^{-9}$   &  2.4$\times$10$^{-10}$  & 1.2$\times$10$^{-10}$    &  
1.3$\times$10$^{-10}$   & 1.4$\times$10$^{-10}$  \\
X(CS)         & 3.6$\times$10$^{-11}$  & 3.4$\times$10$^{-11}$      &  3.3$\times$10$^{-10}$  & $<$8.5$\times$10$^{-11}$ &                         & 
              &            \\
X(C$_2$H)     & $<$2.1$\times$10$^{-11}$ & $<$1.1$\times$10$^{-11}$ &  1.1$\times$10$^{-8}$   &                          &                         &  
              &               \\
X(H$_2$CO)    & 4.5$\times$10$^{-11}$  & 5.3$\times$10$^{-11}$      &  5.0$\times$10$^{-10}$  &                          & $<$1.4$\times$10$^{-12}$  &  
1.0$\times$10$^{-11}$ & $<$1.4$\times$10$^{-11}$ \\
X(SO)         & 4.4$\times$10$^{-11}$  & 4.6$\times$10$^{-11}$      &  $<$3.0$\times$10$^{-11}$   & $<$2.0$\times$10$^{-11}$  &  $<$4.1$\times$10$^{-11}$ &
                        &       \\ 
X(HCN)        &  1.7$\times$10$^{-11}$      &     1.0$\times$10$^{-11}$     &   4.9$\times$10$^{-10}$      &   3.1$\times$10$^{-11}$   &
 1.6$\times$10$^{-11}$  & $<$9.1$\times$10$^{-12}$ & $<$1.1$\times$10$^{-11}$ \\
X(c-C$_3$H$_2$) &  $<$2.4$\times$10$^{-11}$   &     $<$1.0$\times$10$^{-11}$  &   $<$1.3$\times$10$^{-11}$   &   $<$1.3$\times$10$^{-11}$  & & &   \\  
X(DCO$^+$)    &   7.1$\times$10$^{-13}$     &    7.9$\times$10$^{-13}$      &                              &   $<$7.9$\times$10$^{-13}$  & 7.8$\times$10$^{-13}$ & &   \\
X(DCN)        &  $<$3.8$\times$10$^{-12}$   &    $<$4.3$\times$10$^{-12}$   &                              &                            & &  &  \\
X(SiO)        &  $<$1.7$\times$10$^{-11}$   &   $<$2.4$\times$10$^{-11}$    &                              &                            & &  & \\
X(HCO)        &  $<$1.9$\times$10$^{-10}$   &   $<$3.9$\times$10$^{-10}$    &                              &                            & &  &  \\
X(H$_2$S)     &  8.3$\times$10$^{-11}$      &   6.3$\times$10$^{-11}$       &                              &                            & &   &  \\  \hline
X(CN)/X(HCN)          &  6   &  14  &  6    &  8  & 7       &  $>$14  &   $>$13 \\ 
X(CS)/X(SO)           &  0.8 &  0.7 & $>$11 &     &         &         &          \\
X(DCO$^+$)/X(HCO$^+$) &  0.04 & 0.03 &      &     & 0.03    &  0.004       &          \\
X(H$_2$CO)/X(HCO$^+$) &  2.5  &  2   & 0.7  &     & $<$0.06 &  1.3    &  $<$0.14 \\
X(C$_2$H)/X(HCO$^+$)  & $<$1  & $<$0.4 &  15  &   &         &         &          \\
\hline \hline
\end{tabular}

\noindent
$^1$ Abundances relative to H$_2$ have been calculated assuming a disk diameters of 13$"$ for AB~Aur (Pi\'etu et al. 2005), 6$"$ for LkCa15 (Thi et al. 2004) and 13$"$ for DM~Tau (Pi\'etu et al. 2007). The assumed 
disk-averaged molecular hydrogen column densities are: N(H$_2$)=1.1$\times$10$^{22}$~cm$^{-2}$ for AB~Aur, 2.7$\times$10$^{22}$~cm$^{-2}$ for DM~Tau and 1.4$\times$10$^{23}$~cm$^{-2}$ for LkCa15. In our calculations for DM~Tau and LkCa15, we adopt T$_{rot}$=10~K. 
Abundances for the other disks/molecules are taken from Dutrey et al. (1997) (for DM~Tau), Thi et al. (2004) and Guilloteau et al. (2006).
\end{table*}

\section{Averaged molecular abundances in the disk}
In the following we derive approximated average column densities in the disks assuming optically thin emission, a uniform
temperature of T$_{k}$=10~K and 20~K, and Local Thermodynamic Equilibrium (LTE). The assumed disk sizes (diameters) are:
13$"$ for AB~Aur (Pi\'etu et al. 2005), 6$"$ for LkCa15 (Thi et al. 2004) and 13$"$ for DM~Tau (Pi\'etu et al. 2007).
In Table 3 we compare the obtained fractional abundances with those derived in other disks following
a similar procedure. The first result is that the molecular abundances measured toward AB~Aur are very similar to
those found toward other disks which reinforce our interpretation of the disk origin for the observed lines.

The HCO$^+$ abundance in AB Aur is similar to those measured in the TT stars LkCa15 and TW Hya, and in the HAe stars
HD~163296 and MWC~480. Only DM Tau presents a significantly larger (a factor of 10) HCO$^+$ abundance which suggests that DM~Tau
is a special case among circumstellar disks. The same remains true for CN and HCN. Both molecules present abundances
of $\sim$10$^{-10}$ (CN) and $\sim$10$^{-11}$ (HCN) in all the disks except DM Tau in which the measured abundances are
a factor of 10 larger. The case of H$_2$CO is a bit different. It is also overabundant in DM~Tau (5$\times$10$^{-10}$)
but it is underabundant in TW~Hya ($<$10$^{-12}$). Our SO detection in AB Aur is the first one in a circumstellar disk and ours, the first
estimate of the SO abundance in a circumstellar disk.
We have unsuccessfully searched for SO toward DM Tau and LkCa15. Our upper limit toward DM Tau shows that in contrast with
the behavior observed in the other species, SO is not overabundant in DM~Tau relative to AB~Aur. The abundance of C$_2$H in AB~Aur is 1000 times lower than in DM~Tau. We have used the C$_2$H 3$\rightarrow$2 line to derive the upper limit to
the C$_2$H abundance. Recent results by Henning et al. (2010) show that the excitation temperature of C$_2$H could be lower 
than 10~K in some disks. In this case, the value of the upper limit would increase. The searched c-C$_3$H$_2$ line has not been detected
in any of the observed disks. The obtained upper limit to the [c-C$_3$H$_2$]/[HCO$^+$] ratio is still consistent with the observational results in PDRs (Fuente et al. 2003).

In Table~3, we also compare some representative column density ratios. These 
ratios are meaningful providing that the molecules arise in the same region that could not be the case. The value of [CN]/[HCN] is fairly 
uniform (within a factor of 2) among the observed disks. 
However, there are large variations, more than an order of magnitude, in the values of [CS]/[SO], [H$_2$CO]/[HCO$^+$] and
[C$_2$H]/[HCO$^+$]. This suggests that these column density ratios are more sensitive to 
the in most cases poorly known disk structure and/or grain properties.
One important parameter in disks is the deuteration
degree. Qi et al. (2008) derived a [DCO$^+$]/[HCO$^+$] ratio of $\sim$0.03 in TW~Hya.
Our tentative detection of DCO$^+$ in AB Aur would imply a [DCO$^+$]/[HCO$^+$] ratio of $\sim$0.03, similar to TW Hya.

\section{Discussion}
In order to guide our interpretation of the observed features
and provide additional support to their disk origin,
we have performed a (preliminary) chemical model adopting the disk and stellar
parameters from Schreyer et al. (2008) and the updated chemical network 
of Ag\'undez et al. (2008). Our aim is to investigate if detectable SO column densities can be produced in
this disk. In Fig.~4 we show the radial
distribution of the vertical column densities of some molecules as calculated at 2.5~Myr (the age of AB Aur). 
In agreement with the observational results of Pi\'etu et al. (2005), the intense 
stellar radiation makes the disk to be moderately warm, with temperatures above 20~K even in the
disk midplane so that volatile molecules such as CO are not
severely depleted on grain surfaces. In our model,
the major gas phase reservoir of sulphur are the CS and SO molecules, with high column densities of SO 
mainly present in the inner R$<$200~AU region of the disk. The SO abundance decrease rapidly with time because of
the adsorption onto the grain surfaces. The youthness of the AB~Aur disk could be key to have higher SO abundance.

\begin{figure}
\includegraphics{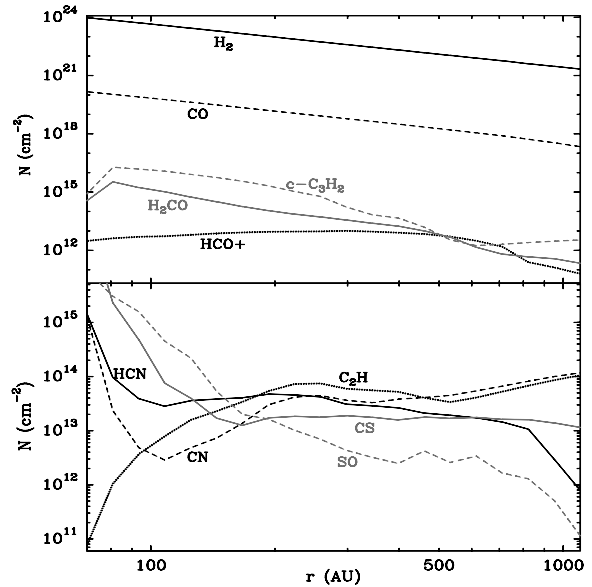}
\vspace{8.0cm}
\caption{Vertical column densities of various molecules as a
function of radius at 2.5~Myr as calculated by the chemical model.} 
\end{figure}

\section{Summary and conclusions}

We have taken advantage of the high sensitivity of the EMIR receivers recently installed in the IRAM
30m telescope to make a sensitive search for molecular emission in three prototypical disks, AB~Aur, DM~Tau and LkCa~15. Our results and conclusions can be summarized as follows:

\begin{itemize}
\item
We have detected the HCO$^+$ 1$\rightarrow$0, CN 1$\rightarrow$0, H$_2$CO 3$_{0,3}$$\rightarrow$2$_{0,2}$, 
SO 3$_{4}$$\rightarrow$2$_{3}$, CS 3$\rightarrow$2, 
HCN 1$\rightarrow$0 and HCN 3$\rightarrow$2 lines toward AB Aur. In addition, we have tentatively detected
the DCO$^+$ 2$\rightarrow$1 and H$_2$S 1$_{1,0}$$\rightarrow$1$_{0,1}$ lines.  Based on the lines profiles, we 
interpret the emission of the CN 1$\rightarrow$0, HCN 3$\rightarrow$2, H$_2$CO 3$_{0,3}$$\rightarrow$2$_{0,2}$, CS 3$\rightarrow$2, and 
the SO 3$_{4}$$\rightarrow$2$_{3}$ lines as arising from the disk. If confirmed, this is the first detection of SO 
in a circumstellar disk. 

\item We have unsuccessfully searched for SO toward DM Tau and LkCa~15. The obtained upper limits show that SO is underabundant 
in DM Tau relative to AB Aur. 

\item We have searched for c-C$_3$H$_2$ toward AB Aur, DM~Tau and LkCa~15. The obtained upper limits are still consistent with the [c-C$_3$H$_2$]/[HCO$^+$]
values obtained in PDRs.
\end{itemize}

Our observational work has significantly increased (from~1~to~6) the number of species detected toward the disk in AB~Aur.
If confirmed by interferometric observations, the SO detection would be the first one in a circumstellar disk.
Our chemical model suggests that the high SO abundance derived in AB Aur disk
is consistent with that expected in a very young and warm disk where
depletion of gas onto grains is not dominating the chemistry yet.

\begin{acknowledgements}
This paper has been partially supported by  MICINN under grant AYA2009-07304 and within the program CONSOLIDER INGENIO 2010, under grant  "Molecular  
Astrophysics: The Herschel and ALMA Era -- ASTROMOL" ( ref.: CSD2009-00038). M.A. is supported by a Marie Curie Intra-European 
Individual Fellowship within the European Community 7th Framework Programme under grant agreement no. 235753. JRG was supported by a Ram\'on y Cajal research
contract from the Spanish MICINN and co-financed by the European Social Fund.

\end{acknowledgements}

\end{document}